\documentclass[12pt]{article}

\def\title{  }

\long\def\abstract{
We show that the F-theory dual of the heterotic string with unbroken
Spin(32)$/\IZ_2$ symmetry in eight dimensions can be described in terms of the
same polyhedron that can also encode unbroken $E_8\times E_8$ symmetry.
By considering particular compactifications with this $K3$ surface as a fiber,
we can reproduce the recently found `record gauge group' in six
dimensions and obtain a new `record gauge group' in four dimensions.
Our observations relate to the toric diagram for the intersection of 
components of degenerate fibers and
our definition of these objects, which we call `tops', is more general than
an earlier definition by Candelas and Font.

}

\def\CY{Calabi--Yau}
\def\ipo{\hbox{\bf 0}}

\input epsf

\def\ifundefined#1{\expandafter\ifx\csname#1\endcsname\relax}
\def\bye{\end{document}}
\long\def\new#1\endnew{{\bf #1}}

\def\HS#1 {\hspace*{#1pt}} \def\VS#1 {\vspace*{#1pt}} \long\def\del#1\enddel{}
\def\BC{\begin{center}}    
\def\EC{\end{center}}

\let\0=\over      \let\1=\vec      \def\2{{1\over2}}    \let\3=\ss
\let\4=\underline \let\5=\overline \let\6=\partial      \def\7#1{{#1}\llap{/}}
\def\8#1{{\textstyle{#1}}}         \def\9#1{{\ifmmode{\pmb{#1}}\else\bf#1\fi}}

          \def\({\left(}       \def\){\right)}

\def\eeql#1 {\label{#1}\eeq}      \let\nn=\nonumber
\def\beq{\begin{equation}}      \def\eeq{\end{equation}}
\def\bea{\begin{eqnarray}}      \def\eea{\end{eqnarray}}

\let\and=\wedge

\let\bra=\langle        \let\ket=\rangle        \def\<#1\>{\bra #1 \ket}
\let\ni=\noindent

\def\rel#1 #2{\buildrel #1 \over {#2}}
\def\fnote#1#2{\begingroup\def\thefootnote{#1}\footnote{#2}
                \addtocounter{footnote}{-1}\endgroup}

\def\subdef#1{\gdef\globalColor##1{##1}}      
      \subdef{Black}

         \let\th=\theta

          \let\G=\Gamma   \let\D=\Delta

\def\IR{\relax{\rm I\kern-.18em R}}
\def\IC{\relax\leavevmode\hbox{\,$\inbar\kern-.3em{\rm C}$}}
\def\IP{\relax{\rm I\kern-.18em P}}
\def\IQ{\relax\leavevmode\hbox{\,$\inbar\kern-.3em{\rm Q}$}}
\def\IK{\relax{\rm I\kern-.18em K}}
\def\II{\relax{\rm I\kern-.18em I}}
\def\IN\def\IN{\relax{\rm I\kern-.18em N}}
\def\IZ{\relax{\sf Z\kern-.4em Z}}
\def\IF{\relax{\rm I\kern-.18em F}}

\def\plb#1 #2 {Phys. Lett. {\bf B#1} #2 }
\def\phr#1 #2 {Phys. Rep. {\bf  #1} #2 }        
\def\npb#1 #2 {Nucl. Phys. {\bf B#1} #2 }
\def\aph#1 #2 {Ann. Phys. {\bf #1} #2 }         
\def\jmp#1 #2 {J. Math. Phys. {\bf #1} #2 }
\def\jgp#1 #2 {J. Geom. Phys. {\bf #1} #2 }
\def\prd#1 #2 {Phys. Rev. {\bf D#1} #2 }
\def\prl#1 #2 {Phys. Rev. Lett. {\bf #1} #2 }
\def\rmp#1 #2 {Rev. Mod. Phys.  {\bf #1} #2 }
\def\zpc#1 {Z. Phys. {\bf #1C} }
\def\cmp#1 #2 {Commun. Math. Phys. {\bf #1} #2 }
\def\cqg#1 #2 {Class.Quant.Grav. {\bf #1} #2 }
\def\mpl#1 {Mod. Phys. Lett. {\bf A#1} }
\def\cpc#1 {Computer Phys. Commun. {\bf #1} }   
\def\ijmp#1 {Int. J. Mod. Phys. {\bf A#1} }
\def\ijmpC#1 {Int. J. Mod. Phys. {\bf C#1} }

\def\BP{\begin{picture}} \def\EP{\end{picture}}         
\newcounter{TRefNX} \let\OLDcite=\cite  \makeatletter
\def\makeTRefs#1{\@for  \NewTRef:=#1\do{\global\makeTRef{\NewTRef}}}
\def\makeTRef#1{\ifundefined{TRef#1}\stepcounter{TRefNX}%
\expandafter\xdef\csname TRef#1\endcsname{\theTRefNX}\fi}\makeatother
\def\NEWcite#1{\makeTRefs{#1}\OLDcite{#1}}

\ifundefined{draftmode} {} \else        \let\cite=\NEWcite
\newcount\HOUR  \HOUR=\the\time \divide\HOUR by 60      \multiply \HOUR by 60
\newcount\MIN   \MIN=\the\time  \advance\MIN by -\HOUR  \divide\HOUR by 60
\ifnum  \MIN>9  \def\printTIME{{\it\the\HOUR\,:\,\the\MIN}}
        \else   \def\printTIME{{\it\the\HOUR\,:\,0\the\MIN}} \fi 

\def\LLab#1{\BP(0,0)\unitlength=1mm\put(-12,.5){\makebox(0,0)[cr]{\small #1
        \rlap{$_{_{\makeatletter\csname TRef#1\endcsname\makeatother}}$}}}\EP}
\fi

\begin{document}
\font\eightrm=cmr8 at 10pt
\let\-=\;

{\eightrm   \hfill UTTG-21-97\vskip -15pt \hfill hep-th/9706226\vskip -15pt 
\hfill June 30, 1997}

\vskip 15mm
\centerline{\huge\bf F-theory, SO(32) and Toric Geometry}

\begin{center} \vskip 10mm
        Philip CANDELAS\fnote{*}{e-mail: candelas@physics.utexas.edu}
\\[5mm]                       and
\\[5mm] Harald SKARKE\fnote{\#}{e-mail: skarke@zerbina.ph.utexas.edu}
\\[3mm] Theory Group\\ Department of Physics\\ University of Texas at Austin\\
        Austin, TX 78712, USA\\[2.5 true cm]
                  {\bf ABSTRACT } \end{center}    
\abstract

\vfill 
\thispagestyle{empty} \newpage
\pagestyle{plain} 

\newpage
\setcounter{page}{1}

\section{\large Introduction}

F-theory \cite{F} was originally defined as a particular D-manifold
vacuum of the type IIB string.
Alternatively, it may be seen as a decompactification of the type IIA
string, thus lifting the well known duality between the type IIA string
and the heterotic string in 6 dimensions \cite{HT,Wstd} to a duality between
the heterotic string on $T^2$ and F-theory on an elliptically fibered
$K3$ surface.
This has afforded new insights into the study of
compactifications of the heterotic string \cite{MVI,MVII}.
The mechanisms by which enhanced gauge groups arise are quite
different on the two sides of the duality:
On the F-theory side, as for the type IIA string, singularities of
the  manifold give rise to gauge groups;
while on the heterotic side, singularities of the manifold
give rise to gauge group enhancement only if small instantons of the
gauge bundle lie precisely on these singularities \cite{Wsi,SW2}.
In ref.~\cite{FMW} several methods for constructing the relevant bundles
were given.
These results were recently used \cite{AM} to derive precisely which gauge
groups and how many tensor multiplets occur when $k$ instantons are placed
on a singularity of type $G$, with $G$  any simply laced Lie 
group.\footnote{Similar results had been derived with different methods
in \cite{Int,BI1,BI2}.}

Practically all  Calabi--Yau manifolds  studied by
physicists have an interpretation in terms of toric geometry, so it is natural
to consider the manifestations of the above phenomena also in these
terms.  Whereas earlier studies in this context \cite{KM,KMP} focused on local
properties, perhaps the most elegant application involves considering  complete
polyhedra: Take an elliptic Calabi-Yau threefold that has $K3$ fibers.
In this case it turns out that the $K3$ polyhedron is contained, as a
subpolyhedron, in the Calabi-Yau polyhedron, and that the Dynkin diagrams 
of the gauge groups that occur upon compactification of the type IIA string 
on the corresponding threefold \cite{CF} can be seen in this $K3$ polyhedron. 
This phenomenon, which can be
explained in terms of the intersection patterns of the toric divisors
\cite{egs},
can of course be lifted to F-theory and has been studied extensively in this
context
\cite{CPR3}. So far, however, these studies have focused on the F-theory or IIA
duals of compactifications of the $E_8\times E_8$ string.
The purpose of the present note is to show that toric methods can equally
easily be used for the description of the duals of the $SO(32)$ string.
In fact, we will find that the same three-dimensional polyhedron that leads
to a $K3$ surface with two $E_8$ singularities may also (upon choosing
a different elliptic fibration and blowing down a different set of toric
divisors) lead to a $D_{16}$ singularity.
This is what we will explain in section two.
In section three we consider two examples of compactifications to lower
dimensions, thereby reproducing the recently found \cite{AM} `record gauge
group' in six dimensions and obtaining a new `record gauge group' in four
dimensions.

\section{\large Eight dimensional vacua}

An eight dimensional F-theory vacuum is determined by an elliptically
fibered $K3$ surface with a section, where all components of the exceptional
fibers except for those intersecting the section are blown down \cite{MVII}.
The gauge group is determined by the $ADE$ classification of quotient
singularities of surfaces.  The intersection matrix of the blown-down
divisors is given by minus the Cartan matrix of the corresponding  Lie algebra.

Many $K3$ surfaces can be constructed as hypersurfaces in toric varieties,
in the following way \cite{Bat}: Consider a pair of reflexive\footnote{ Our
notation is standard.  For a definition of reflexivity and an introduction to
toric geometry written for physicists see, for example, \cite{CDK}.} 
three dimensional polyhedra $(\D,\D^*)$ (for example,
$\D$ could be the Newton polyhedron of a weighted projective space), with
vertices in a pair of dual lattices $(M,N)$.
Then each point in $\D^*$, except for the origin $\ipo$, determines a
ray $v_i$ in $N$.
Each of these rays corresponds \cite{Cox} to a homogeneous
coordinate $z_i$ of the
ambient space $\IP_{\D^*}$ and, consequently, to a divisor $D_i=\{z_i=0\}$.
The $K3$ surface is determined by a divisor in the class $[\sum_i D_i]$.
A divisor in $\IP_{\D^*}$ that does not correspond to a point interior to a
facet (a codimension one face) of $\D^*$ results upon intersection with
the $K3$ surface in a divisor in the $K3$ surface.
Therefore  the intersection pattern of divisors in the $K3$ surface
can be determined by calculating intersections of the type
\beq D_i\cdot D_j\cdot\sum_kD_k   \eeq
in $\IP_{\D^*}$.
This calculation has been considered in detail in \cite{egs}
leading to the following simple rules:
Mutual intersections of divisors in the $K3$ surface are nonzero if and
only if the corresponding points are joined by an edge $\th^*$ of $\D^*$;
in this case their intersection number is the length $l(\th)$ of the
dual edge $\th$ of $\D$ ($\th$ has length $l$ if it has $l-1$ interior
lattice points).
Self-intersections of divisors interior to $\th^*$ are $-2l$; the
corresponding divisors are sums of $l$ rational curves.
The self-intersection of a divisor corresponding to a vertex depends on
the geometry of $\D^*$ in a more sophisticated way.
It is $0$, however, in the case of a regular fiber and $-2$ for any
component of a degenerate fiber of an elliptic fibration.

These results were used to explain the observation made in \cite{CF}
that the extended Dynkin diagrams of enhanced gauge groups are visible
in the toric polyhedron.
As noted in \cite{CF} and explained in \cite{k3}, the fact that a $K3$
surface is an elliptic fibration manifests itself
torically in the fact that $\D^*$ contains the polygon corresponding to
the generic fiber as a reflexive subpolyhedron.

In this section we will consider a particular family of $K3$ surfaces 
which, by the results of \cite{CF}, can give rise to the group 
$E_8\times E_8$; it is the mirror family to the one determined by 
hypersurfaces of degree 12 in $\IP_{1,1,4,6}$.
The principal point that we make in this paper is that the same dual pair of 
polyhedra can also give rise to the gauge group $SO(32)$.
\del
The dual polyhedron, $\D^*$, corresponds to ,  which are mirrors to
those determined by hypersurfaces of degree 12 in
$\IP_{1,1,4,6}$.
\enddel
In suitable coordinates, the dual polyhedron $\D^*$ is the convex hull of the 
points
\beq
(\-1,\-0,\-0),~~~(\-0,\-1,\-0),~~~(-2,-3,\-6),~~~(-2,-3,-6).
\eeq
Altogether $\D^*$ contains 39 lattice points.
By the results of \cite{crp,wtc} there is no three dimensional reflexive
polyhedron with a larger number of lattice points (there is precisely
one other polyhedron with the same number of points, namely the Newton
polyhedron of the degree 6 surface in $\IP_{1,1,1,3}$).
Of these points 21 are `relevant' in the sense that they are not interior to
facets.
As there are three independent relations of linear equivalence among the
divisors in $\IP_{\D^*}$ we see that the Picard lattice of a generic member
of the corresponding family of $K3$ surfaces has rank 18.
Figure 1 consists of a picture of this polyhedron, showing that it
contains two different reflexive triangles.
We have omitted the `irrelevant' points (the points interior to
facets) except those belonging to the reflexive triangles.

\begin{figure}[htb]
\epsfxsize=3.5in
\hfil\epsfbox{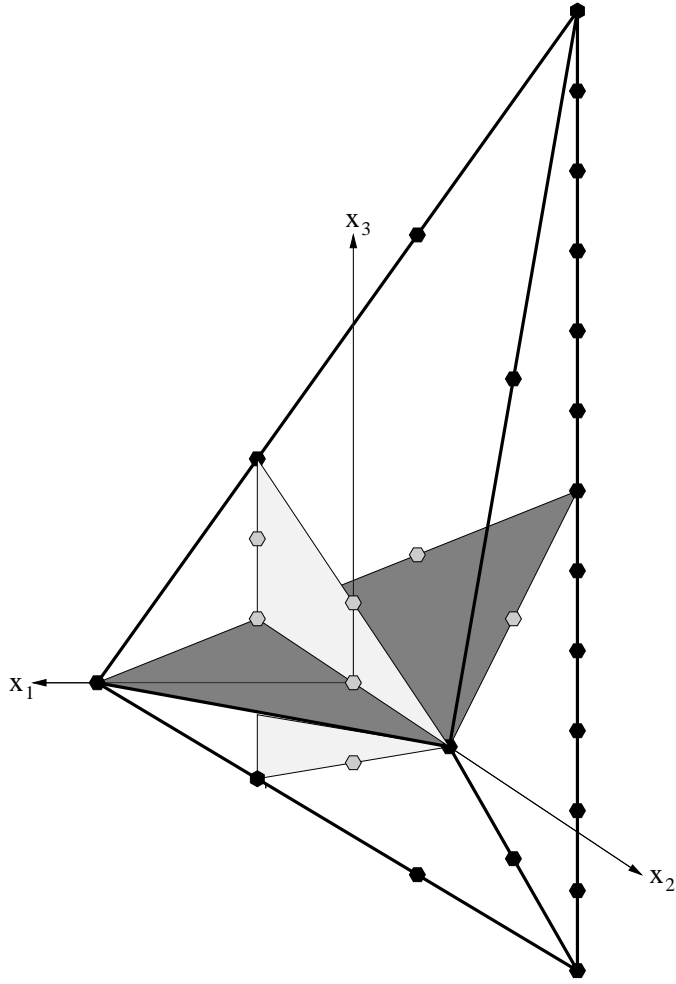}\hfil
\caption{\it The reflexive polyhedron that contains both the Weierstrass and
the new $SO(32)$ triangle.}
\label{fig:poly}
\end{figure}

In the first case this is the well known `Weierstrass triangle',
the convex hull of the points
\beq (\-1,\-0,\-0),~~~(\-0,\-1,\-0),~~~\hbox{and}~~~(-2,-3,\-0), \eeq
lying in the plane $x_3=0$.
The generic fiber is determined by a Weierstrass equation in
the coordinates corresponding to the vertices of this triangle, with
coefficients that depend on the coordinates coming from the points
outside the plane $x_3=0$.
This triangle divides $\D^*$ into two halves, each of which is
an `$E_8$ top' as introduced in \cite{CF}.
The section of the fibration can be identified with the divisor
corresponding to the point $(-2,-3,\-0)$.
The `top' and the `bottom' correspond to extended Dynkin diagrams
of $E_8$, the extension points being the points $(-2,-3,\-1)$ and $(-2,-3,-1)$
above and below the `section point', respectively.

In the second case our triangle lies in the plane $x_1=0$ and is
given as the convex hull of the points
\beq
(\-0,\-1,\-0),~~~(\-0,-1,-2),~~~\hbox{and}~~~(\-0,-1,\-2).
\eeq
This triangle is dual to the Newton polyhedron of $\IP_{1,1,2}$\cite{MVI}.
Now the $K3$ polyhedron is split in an asymmetric way:
On one side we have just a single point, the corresponding divisor being
a smooth fiber, whereas on the other side we have 17 relevant points
forming the extended Dynkin diagram of $SO(32)$.
Here we have two different sections, determined by $(\-0,-1,-2)$ and 
$(\-0,-1,\-2)$, respectively.
This is in agreement with the assertion in \cite{AG2} that a fibration
giving an $SO(32)$ string should have two distinct sections.
Again we take the extension point to be the point adjacent to one of the
`section points'.
Compactification of F-theory on the corresponding $K3$ surface with
the fibers corresponding to the Dynkin diagram blown down should be
dual to the $SO(32)$ heterotic string on a $T^2$ with no Wilson lines on.
It is easily checked by considering the dual polyhedron $\D$ that the complex
structure of this model depends on two free parameters.

The two different fibration structures that can be read off from our
polyhedron $\D^*$ are  related to the fact \cite{AM} that the Picard
lattice $\G_{1,17}$ of the $K3$ 
allows two different decompositions as $\G_{1,1}\oplus\G_{16}$.

\section{\large Lower dimensional compactifications and large gauge groups}

In this section we will give examples of compactifications of F-theory
to six and four dimensions such that the polyhedron of the previous
section encodes K3 fibers of the corresponding compactification
manifolds.
Let us start with considering a Calabi-Yau threefold in the family that is
mirror dual to $\IP_{1,1,12,28,42}[84]$
(this means that $\D^*_{CY}$ will be the Newton polyhedron of
$\IP_{1,1,12,28,42}[84]$).
In suitable coordinates the vertices of $\D^*_{CY}$ are given by
\beq
(\-1,\-0,\-0,\-0),~~~(\-0,\-1,\-0,\-0),~~~(-2,-3,-42,-6),~~~
(-2,-3,\-42,-6),~~~(-2,-3,\-0,\-1).
\eeq
In these coordinates it is easy to see that there are two $K3$ polyhedra
in $\D^*_{CY}$.
One of them lies in the hyperplane  $x_3=0$.
It corresponds to a `trivial top' with an $E_8$ bottom and will not be
considered further.
The other one lies in the hyperplane $x_4=0$.
Its vertices are
\beq (\-1,\-0,\-0,\-0),~~~~ (\-0,\-1,\-0,\-0),~~~~(-2,-3,-6,\-0),~~~~ 
(-2,-3,\-6,\-0).   \eeq
This is just the polyhedron $\D^*_{K3}$ we discussed in the previous section.
The two different elliptic fibration structures in our
polyhedron $\D^*_{K3}$ carry over to $\D^*_{CY}$.
As explained in \cite{fft}, the fan for the toric variety describing
the base of the fibration is determined by projecting the polyhedron
along the fiber (more precisely: by considering equivalence classes
of points differing only by vectors lying in the plane of the fiber).
In the present case, this projection just amounts to throwing
away the second and third coordinate of each point.
Figure 2 shows the image of $\D^*_{CY}$ under this projection and
the resulting fan.

\begin{figure}[htb]
\epsfxsize=6in
\hfil\epsfbox{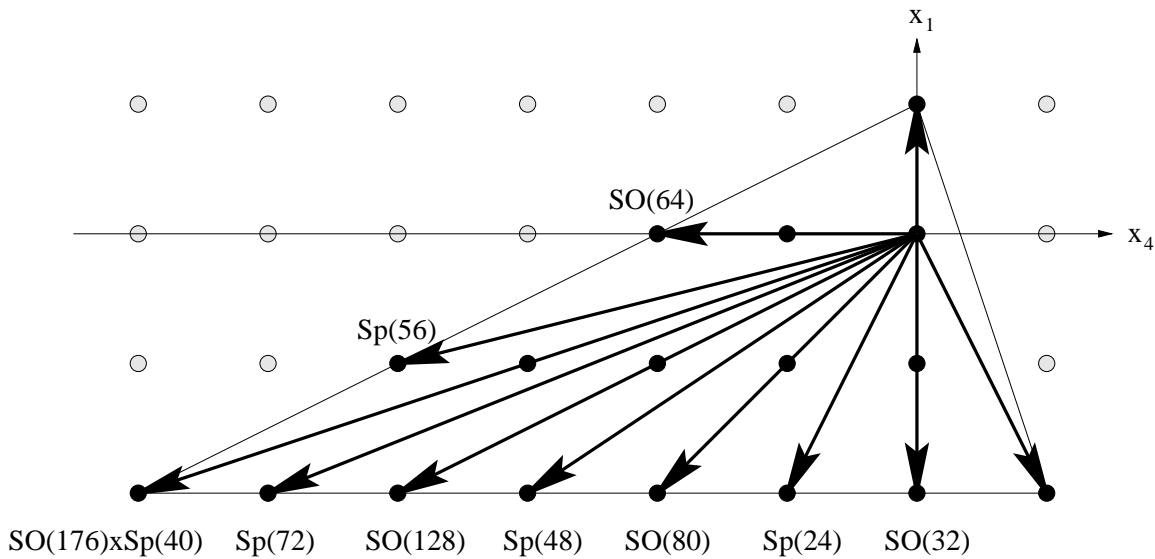}\hfil
\caption{\it The fan of the base of the $SO(32)$-fibration for the threefold.
The rays give rise to the indicated groups.}
\label{base}
\end{figure}

The elliptic fiber can degenerate along the curves in the base space
determined by the toric divisors.
The way it degenerates can be determined by considering the preimage
of the projection.
For the points $(x_1,x_4)=(1,\-0)$ and $(-2,\-1)$ respectively the preimage
consists of only one point, so the corresponding fibers are just
smooth elliptic curves.
The other divisors in the base space lead to enhanced gauge groups as
indicated.
The preimages of the corresponding rays are `tops', where
we define a `top' to be a three dimensional lattice polyhedron with one
facet containing the origin and the other facets at integer distance one 
from the origin. 
This definition implies that the facet containing the origin is a reflexive 
polygon.
The relevant points  of the `top' without the facet containing the origin
form the extended Dynkin diagram of the gauge group.
Note that this generalises the concept of a `top' as a half of a reflexive
polyhedron as formulated in \cite{CF}, since most of the tops we will 
encounter cannot be completed to a reflexive polyhedron.
For the $SO(\cdot)$ groups the `tops' look similar to the right part 
of Figure 1, but with more points along the long line.
As an example for an `$Sp(\cdot)$ top' Figure 3 shows the
`top' for $Sp(24)$.

\begin{figure}[htb]
\epsfxsize=4in
\hfil\epsfbox{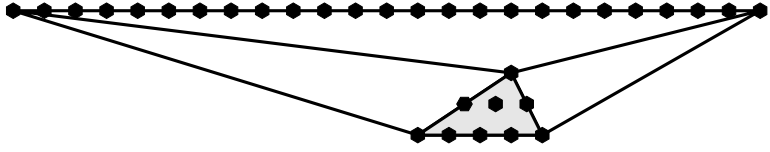}\hfil
\caption{\it The `top' for the group $Sp(24)$. There are 25 points in a 
straight line which form the extended Dynkin diagram of the group.}
\label{sp24}
\end{figure}

There are two types of subtleties that can arise from the fact that
we interpret points as points of $\D^*_{CY}$ on the one hand and as
points of a `top' on the other hand, both related to the fact that
they need not correspond to the same number of divisors in each picture.
The first of these subtleties occurs when a point that corresponds to
precisely one divisor in the Calabi-Yau space corresponds to more
than one divisor in the top.
In this case several divisors that form part of an ADE pattern are
identified globally; the corresponding monodromy leads to a non-simply
laced gauge group.
This is how we get the $Sp(\cdot)$ groups.
The other special case is when points that are irrelevant in the
context of the `top' turn out to be relevant for the Calabi-Yau space, as
discussed in \cite{CPR3}.
This happens for the point $(-2,-6)$ in the base.
The preimage of the corresponding ray looks like the
`top' for $SO(176)$, but after identifying the points that are relevant
in the context of the `top' with the corresponding Dynkin diagram,
we find that there are still points that are relevant in the context of
the whole Calabi-Yau space.
These points, by themselves, form an $Sp(40)$ `top'.
Since in this case the extra points are relevant, the total contribution
from the preimage of this ray is $SO(176)\times Sp(40)$.

Proceeding similarly with what is possibly the `largest' fourfold polyhedron,
the Newton polyhedron of $\IP_{1,1,84,516,1204,1806}[3612]$, we 
may choose coordinates such that the vertices of $\D^*_{\rm fourfold}$ 
are given by
\bea
(\-1,\-0,\-0,\-0,\-0),~~~~ (\-0,\-1,\-0,\-0,\-0),~~~~(-2,-3,\-3606,-6,-6),
~~~~(-2,-3,-6,\-37,-6),\nn\\
\hfill (-2,-3,-6,-6,\-1),~~~~ (-2,-3,-6,-6,-6).\hspace{3.5truecm}
\eea
The $E_8\times E_8$ or $SO(32)$ polyhedron lies in the plane $x_4=x_5=0$ and 
the $SO(32)$ triangle lies in the $x_2x_3$-plane.
The base of the fibration is therefore determined by projecting out $x_2$ 
and $x_3$.
This results in a tetrahedron $T$ with vertices 
\beq
(\-1,\-0,\-0),~~~~ (-2,-6,-6),~~~~ (-2,\-37,-6),~~~~(-2,-6,\-1).
\eeq
The preimages of its vertices are single points, the
exception being the vertex $(-2,-6,-6)$ whose preimage consists of 
$3613$ lattice points in a row. 
Together with the preimage of $(-1,-3,-3)$ which lies on the same ray,
this gives the `top' for the group $SO(7232)$.
The points that are irrelevant in the context of this top are relevant
in the context of the fourfold, so we get an additional group factor
of $Sp(1804)$.
The preimages of other points along the boundary of $T$ are always lines
whose numbers of points can be determined by linearity; e.g. the preimage
of 
\beq (-2,\-0,\-0)={6\0 7}(-2,-6,\-1)+{6\0 43}(-2,37,-6)+
{1\0 7\times 43}(-2,-6,-6)
\eeq
consists of
\beq {6\0 7}\times 1+{6\0 43}\times 1+{1\0 7\times 43}\times 3613=13
\eeq
points giving rise to the unbroken $SO(32)$.
In general $k$ points in a row give rise to $SO(2k+6)$ or $Sp(k-1)$,
depending on whether the point in the boundary of $T$
is divisible by two or not.
If a point that leads to an $SO(\cdot)$ group lies on an edge of $T$,
then there is an additional factor of $Sp({k-5\0 2})$.
Altogether we get a gauge group of rank 302,896
consisting of 251 non-trivial simple factors, the largest
of them being $SO(7232)$ and $Sp(3528)$ and the smallest being $SO(32)$ 
and $Sp(24)$.
The full list of factors is given in the appendix.
It is
interesting that the $SO(32)$ decomposition seems to produce groups of larger
rank than the $E_8\times E_8$ decomposition.  For the fourfold we have been
considering the $E_8\times E_8$ decomposition leads to the group \cite{CPR3} 
$E_8^{2561}\times F_4^{7576}\times G_2^{20168}\times SU(2)^{30200}$  
of rank $121,328$ while for the $SO(32) $ decomposition the rank is larger.
The reason for this is that the number of divisors in the base is smaller
in the $SO(32)$ case because here we project along the direction along
which the extension of the polytope is largest.

\ni
{\bf \large Acknowledgements:}
We would like to thank Eugene Perevalov and Govindan Rajesh for useful
discussions.
This work is supported by the Austrian `Fonds zur
F\"orderung der wissenschaftlichen Forschung' (Schr\"odinger
fellowship J012328-PHY),
by NSF grant PHY-9511632 and the Robert A. Welch Foundation.

\newpage
\ni
\centerline{\bf \large Appendix: The groups that arise from the toric 
divisors of the base}
\vskip 0 true mm
\font\sixrm=cmr6 at 7.2pt
$$
\def\skip{\hskip3.4pt}
\vbox{\offinterlineskip\halign{
\vrule height8pt depth6pt width0pt # 
&\hfil\skip\sixrm  #, &\hfil\skip\sixrm  #, 
&\hfil\skip\sixrm  #, &\hfil\skip\sixrm  #, 
&\hfil\skip\sixrm  #, &\hfil\skip\sixrm  # \cr
&Sp[0]&Sp[168]&Sp[336]&Sp[504]&Sp[72]&Sp[156],\cr
&Sp[672]&Sp[240]&Sp[324]&Sp[840]&Sp[408]&Sp[492],\cr
&Sp[1008]&Sp[56]&Sp[576]&Sp[140]&Sp[144]&Sp[660],\cr
&Sp[1176]&Sp[224]&Sp[744]&Sp[308]&Sp[312]&Sp[828],\cr
&Sp[1344]&Sp[392]&Sp[912]&Sp[476]&Sp[480]&Sp[996],\cr
&Sp[1512]&SO[176]\ Sp[40]&Sp[48]&Sp[560]&Sp[1080]&SO[512]\ Sp[124],\cr
&Sp[132]&Sp[644]&Sp[648]&Sp[1164]&Sp[1680]&SO[848]\ Sp[208],\cr
&Sp[216]&Sp[728]&Sp[1248]&SO[1184]\ Sp[292]&Sp[300]&Sp[812],\cr
&Sp[816]&Sp[1332]&Sp[1848]&SO[1520]\ Sp[376]&Sp[384]&Sp[896],\cr
&Sp[1416]&SO[1856]\ Sp[460]&Sp[468]&Sp[980]&Sp[984]&Sp[1500],\cr
&Sp[2016]&SO[128]&SO[2192]\ Sp[544]&Sp[552]&Sp[1064]&Sp[1584],\cr
&Sp[112]&SO[464]&Sp[120]&SO[2528]\ Sp[628]&Sp[636]&Sp[1148],\cr
&Sp[1152]&Sp[1668]&Sp[2184]&SO[800]&SO[2864]\ Sp[712]&Sp[720],\cr
&Sp[1232]&Sp[1752]&Sp[280]&SO[1136]&Sp[288]&SO[3200]\ Sp[796],\cr
&Sp[804]&Sp[1316]&Sp[1320]&Sp[1836]&Sp[2352]&SO[1472],\cr
&SO[3536]\ Sp[880]&Sp[888]&Sp[1400]&Sp[1920]&Sp[448]&SO[1808],\cr
&Sp[456]&SO[3872]\ Sp[964]&Sp[972]&Sp[1484]&Sp[1488]&Sp[2004],\cr
&Sp[2520]&SO[64]&SO[80]&Sp[24]&SO[2144]&SO[4208]\ Sp[1048],\cr
&Sp[1056]&Sp[1568]&Sp[2088]&SO[400]&SO[416]&Sp[108],\cr
&Sp[616]&SO[2480]&Sp[624]&SO[4544]\ Sp[1132]&Sp[1140]&Sp[1652],\cr
&Sp[1656]&Sp[2172]&Sp[2688]&SO[736]&SO[752]&Sp[192],\cr
&SO[2816]&SO[4880]\ Sp[1216]&Sp[1224]&Sp[1736]&Sp[2256]&SO[1072],\cr
&SO[1088]&Sp[276]&Sp[784]&SO[3152]&Sp[792]&SO[5216]\ Sp[1300],\cr
&Sp[1308]&Sp[1820]&Sp[1824]&Sp[2340]&Sp[2856]&SO[1408],\cr
&SO[1424]&Sp[360]&SO[3488]&SO[5552]\ Sp[1384]&Sp[1392]&Sp[1904],\cr
&Sp[2424]&SO[1744]&SO[1760]&Sp[444]&Sp[952]&SO[3824],\cr
&Sp[960]&SO[5888]\ Sp[1468]&Sp[1476]&Sp[1988]&Sp[1992]&Sp[2508],\cr
&Sp[3024]&Sp[0]&SO[32]&SO[2080]&SO[2096]&Sp[528],\cr
&SO[4160]&SO[6224]\ Sp[1552]&Sp[1560]&Sp[2072]&Sp[2592]&SO[352],\cr
&SO[368]&Sp[96]&SO[2416]\ Sp[600]&SO[2432]&Sp[612]&Sp[1120],\cr
&SO[4496]&Sp[1128]&SO[6560]\ Sp[1636]&Sp[1644]&Sp[2156]&Sp[2160],\cr
&Sp[2676]&Sp[3192]&SO[704]&Sp[688]&SO[2768]&Sp[696],\cr
&SO[4832]&SO[6896]\ Sp[1720]&Sp[1728]&Sp[2240]&Sp[2760]&SO[1040]\ Sp[256],\cr
&Sp[264]&SO[3104]\ Sp[772]&Sp[780]&SO[5168]\ Sp[1288]&Sp[1296]&SO[7232]\ Sp[1804],\cr
&Sp[1812]&Sp[2324]&Sp[2328]&Sp[2844]&Sp[3360]&Sp[344],\cr
&Sp[860]&Sp[864]&Sp[1376]&Sp[1892]&Sp[1896]&Sp[2408],\cr
&Sp[2928]&Sp[432]&Sp[948]&Sp[1464]&Sp[1980]&Sp[2496],\cr
&Sp[3012]&Sp[3528]&Sp[0]&Sp[1032]&Sp[2064]&Sp[3096].\cr
}}
$$

\bye